\documentclass[
aps,%
12pt,%
final,%
notitlepage,%
oneside,%
onecolumn,%
nobibnotes,%
nofootinbib,% 
superscriptaddress,%
noshowpacs,%
centertags]%
{revtex4}

\begin{document}
\selectlanguage{russian}

\title{Detection of Fast Radio Bursts on the Large Scanning Antenna of the Lebedev Physical Institute}

\author{\firstname{V.~€.}~\surname{Fedorova}}Fedorova V.€.
\email{fedorova-astrofis@mail.ru}
\affiliation{Pushchino Radio Astronomy Observatory, Astro Space Center, Lebedev Physical Institute, Pushchino, Moscow region, Russia}
\author{\firstname{€.~….}~\surname{Rodin}}Rodin €.…. 
\email{rodin@prao.ru}
\affiliation{Pushchino Radio Astronomy Observatory, Astro Space Center, Lebedev Physical Institute, Pushchino, Moscow region, Russia}

%\date{\today}
%\today ¯¥ç â ¥â c¥£®¤­ïè­¥¥ ç¨á«®

\begin{abstract}
{\bf{Abstract}} -- Results of a search for individual impulsive signals on the Large Scanning Antenna of the Lebedev Physical Institute at 111 MHz carried out from July 2012 through May 2018 are presented. The data were convolved with a template of a specified form and convolved with a test dispersion measure. A region of sky with central coordinates $\alpha = 05^h 32^m;$ $\delta = +41.72^\circ$ and also a region of sky around the coordinates fixed earlier for FRB 121102 ($\alpha = 05^h 32^m;$ $\delta = +33.1^\circ$) were chosen for the analysis. In all, 355 hours of observations were processed for each beam. Three radio bursts with dispersion measures of $247$ $pc \cdot cm^{-3}$, $570$ $ pc \cdot cm^{-3}$, $1767$ $ pc \cdot cm^{-3}$ were detected in the course of reducing the data.

\end{abstract}

\maketitle

\section{INTRODUCTION}

Searches for fast radio bursts (FRBs) are one of the most topical areas of study in modern astrophysics, due in good part to the mysterious nature of this phenomenon. The first FRB was detected in 2007 in the archives of the Parkes radio telescope containing data on pulsar searches \cite{lorimer}. Doubt was cast on the natural origin of these and some other bursts \cite{champion, spitler}, since the associated signals were isolated and had
parameters similar to peritones \cite{burke-spolaor}. After the detection of a repeat signal from FRB 121102 \cite{scholz}, the possibility of an artificial origin for these signals was eliminated, and the central question became what processes could give rise to such powerful pulses.

Several mechanisms for the origin of radio bursts such as FRBs have been proposed \cite{wei-min, bramante}. FRB 121102 is an event that has interested many researchers. In connection with the detection of multiple repeat events occurring over several years, it has been possible to establish the coordinates from which the impulsive signal is coming with fairly good accuracy. This region was also investigated in the optical using the Hubble Space Telescope and the 8.2-m Subaru telescope. As a result, it was determined that FRB 121102 is located in a region of active star formation very near the center of an irregular dwarf galaxy at a distance of more than three billion light years \cite{KOKUBO}.

Much attention is being paid to theoretical models
that can satisfactorily explain all the parameters of
FRBs. Some 20 or more models are known in the
literature, but the most commonly considered models
are those where FRBs are gigantic pulsar pulses \cite{popov}.

More than 50 FRBs are currently known. These all have dispersion measures from $pc/cm^3$ to 2596.1 $pc/cm^3$. This suggests that most of these
signals have an extragalactic origin. The pulse
durations for all these events do not exceed several
milliseconds. So far, all the FRB detections have been
at frequencies from 700 MHz to 8 GHz.

One of the important characteristics of FRBs is
their brightness: the peak fluxes for the vast majority
of these bursts lie in the range 0.3 -- 8 Jy. In spite
of the similarity of the parameters of FRBs, we can
distinguish several special cases. Three bursts having peak fluxes several times these values have been
detected. For example, the peak flux of FRB 150807
was 120 Jy \cite{ravi}.

Circular and linear polarization was detected for FRB 110523, FRB 140514, and FRB 150807 \cite{ravi}. Repeated pulses from the direction of FRB 121102 were detected on three occasions \cite{scholz, chatterjee}. FRB 180411 with a record signal-to-noise $SNR$ = 411 was detected in April 2018 \cite{Oslowski}. This pulse was detected using the Parkes 64-m radio telescope (Australia) at 1.4 GHz.

The main goal of our current study is to analyze data obtained with the Large Scanning Antenna (LSA) of the Lebedev Physical Institute during routine observations of radio sources scintillating on imhomogeneities of the interplanetary plasma, in order to search for impulsive signals with properties similar to FRBs.

The following sections describe the technical characteristics of the LSA, consider numerical simulations of the detection of a signal of a specified form against a background of white noise, and present results for LSA data obtained from 2012 through 2018.

\section{APPARATUS}

The LSA is a meridian instrument, and is one of the most sensitive radio telescopes operating at meter wavelengths. Starting in 2000, the operating frequency range of this instrument has been 111 ŒHz $\pm$  1.25 ŒHz. The fluctuational sensitivity of the radio telescope in a 2.5-MHz receiver bandwidth with a time resolution of 0.1 s is 140 mJy \cite{oreshko}. The signal is registered using a multichannel digital receiver that enables the signal to be recorded in two modes. The first recording mode has relatively low frequency resolution and six frequency channels, each with a receiver bandwidth of 415 kHz. The time interval between readouts is 100 ms. The data obtained in this mode are used to continuously monitor scintillating sources. The second recording mode uses 32 frequency channels with a bandwidth of 78 kHz and a time resolution of 12.5 ms. In both the first and the second mode, the signals are digitally processed using a 512-sample FFT processor. In
our current study, we used data recorded with a time resolution of 100 ms.

To study large numbers of interplanetary scintillations of compact radio sources in a monitoring regime, a stationary, 96--beam antenna beam was created, which covers the sky from $-9^{\circ}$  to $42^{\circ}$  in declination. The full width at half-maximum of an
individual beam depends on the declination of the observed source, and ranges from $24^\prime$ to $48^\prime$. The time for the passage of a source through the antenna
beam is 4--7 min. The maximum effective area of the antenna (47 000 m$^2$) is realized at the zenith, and decreases toward the horizon as $\cos z$, where $z$ is the zenith angle. The system noise temperature fluctuates in the range 550--3500 K, depending on the sky background. An advantage of the LSA for searching for signals similar to FRBs is its large field of view ($\backsim$ 50 square degrees), and also its continuous and round-the-clock monitoring of the sky, with recording of data on a server for subsequent reduction.

\section{MATHEMATICAL MODELING}

Our simulations of the detection of an impulsive signal included several steps. Since an impulsive signal undergoes scattering on inhomogeneities in the interstellar medium, the shape of the received signal differs from the shape of the emitted pulse. In our case, the parameters of our model for scattering of an impulsive signal depended on the dispersion measure $DM$,  the central receiver frequency, and the channel bandwidth. The dependence of the pulse scattering on the dispersion measure $DM$ at 110 MHz, $t_s=0.06(\frac{DM}{100})^{2.2}$, was taken from Kuzmin et al.\cite{kuz}. To model a signal, we generated a sequence of 3000 points with a sampling interval of $\Delta t =$ 0.1 swhich on average corresponds to a time for passage of the observed source through the LSA antenna beam of 5 min.

If we adopt a thin-screen model \cite{uil}, the impulsive characteristics of the scattering medium can be described by the decaying exponential function

\begin{equation}
h = \frac{1}{t_s} e^{-{\frac{t}{t_s}}}.
\end{equation}

The received pulse is the convolution of the emitted pulse, represented by a delta-function, and the transfer function of the medium \cite{uil}

\begin{equation}
F(t) = \frac 1 {t_s} \int {e^{-{\frac{t}t_s}}} \delta (t - \tau) d\tau,
\end{equation}
where $t$ $-$ is the delayed time for the arrival of the signal at an individual frequency channel. As an example, Fig. 1 presents the function $F(t)$, representing the scattered pulse arriving at each of six frequency channels. We used the dispersion measure $DM = 360$ pc$\cdot$cm$^{-3}$, in these simulations, which yields $t_s = 1$ s.

The received pulse, which, in addition to scattering, has been subject to a dispersion delay in the medium through which it propagates, is received in the finite frequency band, leading to broadening, which can also be described as a convolution of the received pulse $F(t)$ with a $\Pi$-- shaped function represented as a product of two Heaviside (unit jump) functions $\sigma(t)$:

\begin{equation}
\Pi (t) = \sigma(t - \tau_{i-1})\sigma(\tau_i - t)
\end{equation}
where $\tau_i$ is the arrival time at the boundary frequency of frequency channel $i$. The quantity $\Delta\tau=\tau_i -\tau_{i-1}, (i = 1,2,3,\dots, 6)$ represents broadening of the
pulse in the band. Figure 2 shows pulses distorted by broadening in the frequency channels and shifted by the dispersion delay. We note especially that the broadening in the bandwidth $\Delta\tau$ and scattering $t_s$ depend on $DM$ in completely different ways: $\sim DM$ and $\sim DM^{2.2}$, respectively, $\Delta\tau=t_s$ with $DM \approx 320$ pc$\cdot$/m$^3$. Therefore, when $DM \lesssim 320$ pc$\cdot$/m$^3$,the main contribution to smearing of the
pulse is made by broadening in the bandwidth, while the main contribution when $DM \gtrsim 320$ pc$\cdot$/m$^3$ is made by scattering in the intervening medium.
 
In the next step, six realizations of additive white noise with various amplitudes were superposed on the pulse broadened and subject to dispersion delay. We specified various SNR levels from 0.1 to 5, in accordance with real observations. The rms deviations of the additive noise were calculated as a function of the amplitude of the dispersed pulse and the specified SNR. Figure 3 shows an example of noisy impulsive signals in the six frequency channels.

Because the data reduction occupies a fairly long time, we compared several methods for distinguishing an impulsive signal from the point of view of the duration of the reduction and efficiency of the signal detection. We applied three approaches to distinguishing a dispersed impulsive signal: addition of pairwise crosscorrelation of the pulses in the frequency channels, and cross - correlation of a noisy pulse with followed by addition of these pulses, cross - correlation with a template followed by addition and compensation for the dispersion measure (Fig. 4). Figure 4 shows that, for the same input noise level, this last method based on cross - correlation with a template enables identification of the signal with the maximum SNR.
We therefore decided to use this method, in spite of the fact that it required the longest processing time.

\section{DATA REDUCTION}

The data reduction was carried out in several steps. First, corrections were introducted to compensate for small deviations of the plane of the LSA beams from the local meridian, and also for precession. Further, a background smoothed with a median filter was subtracted from recordings with a duration of one hour.

To search for FRBs, the recordings for the six frequency channels were processed with a time resolution of 0.1 s. Daily data in two of the 96 beams obtained from July 2012 through May 2018 were analyzed. In one case, the recordings of the second LSA beam were used ($\delta = +41.72^\circ$). A section
with the central coordinates $\alpha = 05^h 32^m$ $\pm10^m$; $\delta = +41.72^\circ$was selected from the hourly recordings. In the other case, a section of data obtained in beam 22 ($\delta = +33.25^\circ$) corresponding to the coordinates of the known object FRB 121102 with $\alpha = 05^h 32^m$ $\pm$ $10^m$; $\delta = +33.25^\circ$ was analyzed. We analyzed 355 hours of recordings in each beam in this way.

To verify the correct operation of the program used to identify FRBs, the entire reduction algorithm was
tested using two versions of pulses: a model impulsive signal with $DM = 360$ $pc \cdot{ám^{-3}}$ and $DM = 2000$ $pc \cdot{ám^{-3}}$ was superposed on an hourly recording of LSA data (Figs. 5a and 5b). Further testing was carried out using a real object -- the pulsar B2154+40 (Figs. 5c and 5d). In all cases, the method used enabled reliable identification of the impulsive signal.

In the next step, we analyzed a five-minute section with the central coordinates indicated above. Each
recording was first convolved with a template obtained from our simulations for $DM = 360$ $pc \cdot{ám^{-3}}$. We did not initially carry out convolutions with templates corresponding to different $t_s$ values, since this would have dramatically increased the reduction time. This operation was carried out after the detection of the pulses, to refine the scattering values $t_s$ presented in the table. We then carried out a convolution with a test dispersion measure in the range from 0 ¤® 3000 $pc \cdot{ám^{-3}}$ in steps of 50. Further, we conducted a visual analysis of the results obtained for the convolution with a test $DM$, consisting of searching for a dispersed signal in the recordings convolved with the template, and the presence of a signal with a high amplitude in the data convolved with the test dispersion measure. We then refined the dispersion measures for any signals found and constructed the corresponding integrated pulses.

We obtained more precise estimates of the flux densities of the FRBs that were found by comparing the signals received with a calibration step and calibration source. The calibration step is a template noise signal that is added to the recordings every four hours. We used 3C 48 as a calibration source, since it is located in the same beam as FRB 121102. As a result of determining the flux density of 3C 48, the fluxes were obtained with uncertainties not exceeding 10$\%$.

We did not take into account the possibility that pulses fell between the LSA beams, or did not arrive
at the center of an LSA beam used. Therefore, the peak flux densities presented in the table should be
taken to be lower limits.

\section{RESULTS}

Our visual analysis enabled us to identify three events with $DM$ values $247$ $pc \cdot ám^{-3}$, 570 $pc \cdot ám^{-3}$ ¨ 1767 $pc \cdot ám^{-3}$ (¨áã­ª¨ 6--8) \cite{rodin}.

Since our FRB search was carried out at low frequencies, pulses with high dispersion measures have undergone appreciable broadening. We also assumed that the peak flux densities of the bursts at the observed frequency were essentially at the sensitivity limit of the LSA. This means that it is not possible to directly detect these signals under these conditions. Accordingly, we had to carry out a convolution with a template with a specified form. As a result, the profiles of each of the pulses in Fig. 7 have essentially the same widths. We should also note that we can register only the upper part of a pulse using this approach, since the exponential "tail"$ $ of the signal cannot be distinguished in
the noise.

Based on the model YMW16 (see \cite{model}), we estimated the redshifts $z$ for all three pulses, determined using the relation
\begin{equation}
z = \frac{DM_{IGM}H_0}{c\cdot n_{IGM}}=\frac{[DM-(DM_{Gal}+DM_{MC}+DM_{Host})]H_0}{c \cdot n_{IGM}},
\end{equation}
where $DM$ is the dispersion measure of the observed FRB, $DM_{Gal}$ the total dispersion meaure along the line of sight toward the FRB, $DM_{MC}$ he dispersion of the Magellanic Cloud, $DM_{IGM}$ the dispersion measure of the intergalactic medium, $DM_{Host}$ the dispersion measure of the host galaxy, $H_0$ the Hubble constant ($H_0 = 67.3$ $ª¬ \cdot á^{-1}$), and $n_{IGM}$ the electron number density ($n_{IGM} = 0.16$ $m^{-3}$).

We also estimated the SNR and the peak flux density of each impulse. The right ascension of the impulses
corresponds to the sixth frequency channel with $f = 111.5$ MHz. The results of all our estimates and
computations are presented in the table.

\section{DISCUSSION}

All the detected impulses with measured dispersion measures and peak flux densities have parameters similar to those of previously discovered FRBs. Accordingly, we suggest that all three signals are new
FRBs first detected at this low frequency. The spectral indices of these events are an urgent question.
It has been proposed in various studies that the flux density grows toward higher frequencies. However,
the measured flux densities suggest that the spectrum of the detected FRBs has a spectral index around
zero, or rises slightly toward low frequencies. The detection of FRBs with the LSA at 111 MHz supports
this picture.

Since the discovery of the first FRB and continuing up to the present time, theoretical studies have put
forward a large number of models describing possible mechanisms for the generation of powerful impulses
in the form of FRBs at cosmological distances \cite{popov,div, Yu, houd}. Several mechanisms for the formation of FRBs suggest they should be detected only at high frequencies, with the signal being so weak at low frequencies that it should be essentially impossible to detect \cite{YueZhang, Long-Biao, bah}. Our discovery of FRBs at low frequency places constraints on models that predict a flux that rises
toward higher frequencies. Such models include, for example, an overflow of matter from an accretion disk around a comapct object in a close binary system, the coalescence of charged black holes, and the coalescence of white dwarfs. These constrains can also be applied to scenarios in which FRBs are afterglows of gravitational -- wave events.

One of the main goals of our study was to try to detect impulsive emission from the known object FRB 121102. Our reduction of data in an area with central coordinates $\alpha = 05^h 32^m$; $\delta = +33.1^\circ$ enabled the detection of a pulse with dispersion measure 570 $pc/ám^3$. This agrees with the dispersion measure of FRB 121102 within the uncertainties. The two components of the pulse must be distinguished through a separate analysis. Since there have been multiple attempts to detect several pulses from FRB 121102 within a single observing session, the double profile of the signal is consistent with the nature of this repeating burst.

\section{CONCLUSION}

We will now summarize the main results of this study.
\begin{enumerate}
\item We have developed an algorithm that can be used to search for fast radio bursts at meter wavelengths with the Large Scanning Antenna.
\item Data for an area of sky coincident with the known object FRB 121102 have been reduced: 355
hours of observations in an area with the central coordinates $\alpha = 05^h 32^m; \delta = +33.1^\circ$ yielded the detection of a signal with $DM = 570$ $pc/ám^3$, which agrees within the uncertaities with the dispersion measure for impulses from FRB 121102, which varies from $555$ $pc/ám^3$ to $568.8$ $pc/ám^3$ according to data from an FRB Catalogue. The table shows that the right ascension $\alpha$ for our detected impulse differs from the right ascension of FRB 121102 by $10^m$. This disrepancy could have two possible origins. One is that we have repeatedly detected a signal from
FRB 121102 in one of the side lobes. In this case, if the impulse had been detected in the main beam of
the LSA, its peak flux density would have been appreciably higher, again underscoring our hypothesis
concerning the spectral indices of FRBs. The other possibility is that we have detected an impulse from a completely new FRB source.
\item In any case, we have detected three FRBs with dispersion measures of 247 $¯ª/á¬^3$, 570 $¯ª/á¬^3$, and 1767 $¯ª/á¬^3$. in the interval from July 2012 through May 2018. The parameters of these impulsive events are presented in the table.
\end{enumerate}

ACKNOWLEDGMENTS

This study was supported by the Russian Foundation for Basic Research (project 16--29--13074).

\appendix

\newpage
\begin{center}
{\bf Table 1.} Parameters of the detected impulses
\end{center}

\begin{tabular*}{\linewidth}{|p{2cm}|p{2.3cm}|p{2cm}|p{1.3cm}|p{2cm}|p{1.8cm}|p{1.9cm}|p{1.9cm}|} \hline
Date & Coordinates (J2000), $\alpha, \delta$ & Dispersion measure, $pc/ám^3$ &  SNR & Peak flux density,
Jy & Energy, Jy $\cdot$ ms& Scattering $t_s$, s & Redshift $z$ \\ \hline
18.10.2015 & 0521 +33.1 & 570 $\pm$ 5 &  6.2 & 1.4 & 3500 & 0.275 & 0.273 \\ \hline
20.09.2016 & 0534 +41.7 & 1767 $\pm$ 4 & 9.1 & 0.22 & 1100 & 4.33 & 1.973 \\ \hline
06.06.2017 & 0534 +41.7 & 247 $\pm$ 4 & 8.3 & 0.54 & 1890 & 0.275 & 0.057\\ \hline
\end{tabular*}

\newpage
\begin{figure}[h!]
	\setcaptionmargin{1mm}
	\vbox{\includegraphics[width=0.75\linewidth]{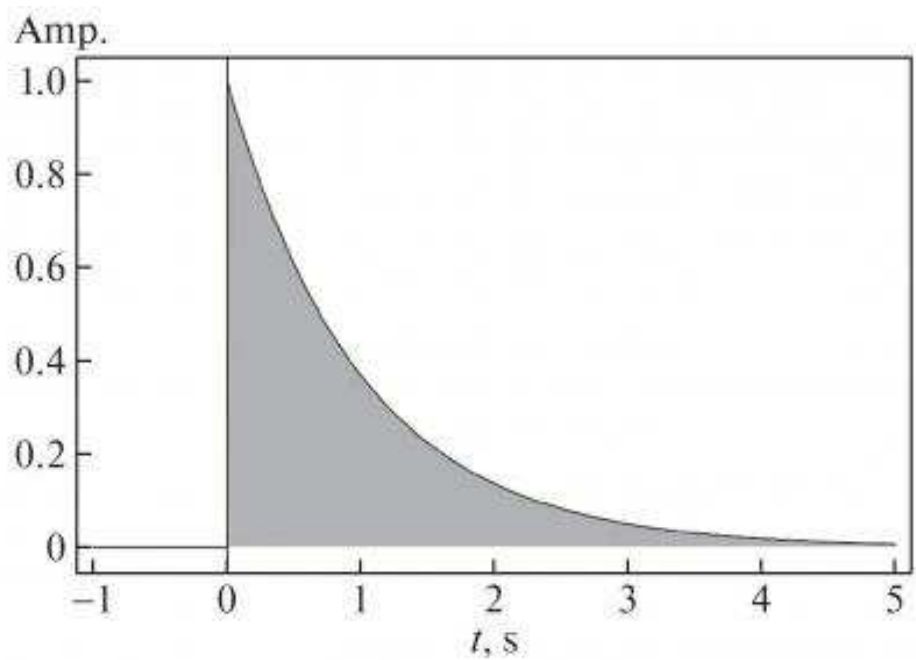}}
	\caption{Simulations of the arrival of an impulsive signal at each of the six frequency channels of the LSA. The scattering of the impulse is $t_s$ = 1 s, which corresponds
	to $DM$ = 360 $pc/ám^3$.}
	\label{ris:fig1}
\end{figure}

\newpage
\begin{figure}[h!]
\begin{minipage}[h]{0.49\linewidth}
\center{\includegraphics[width=1\linewidth]{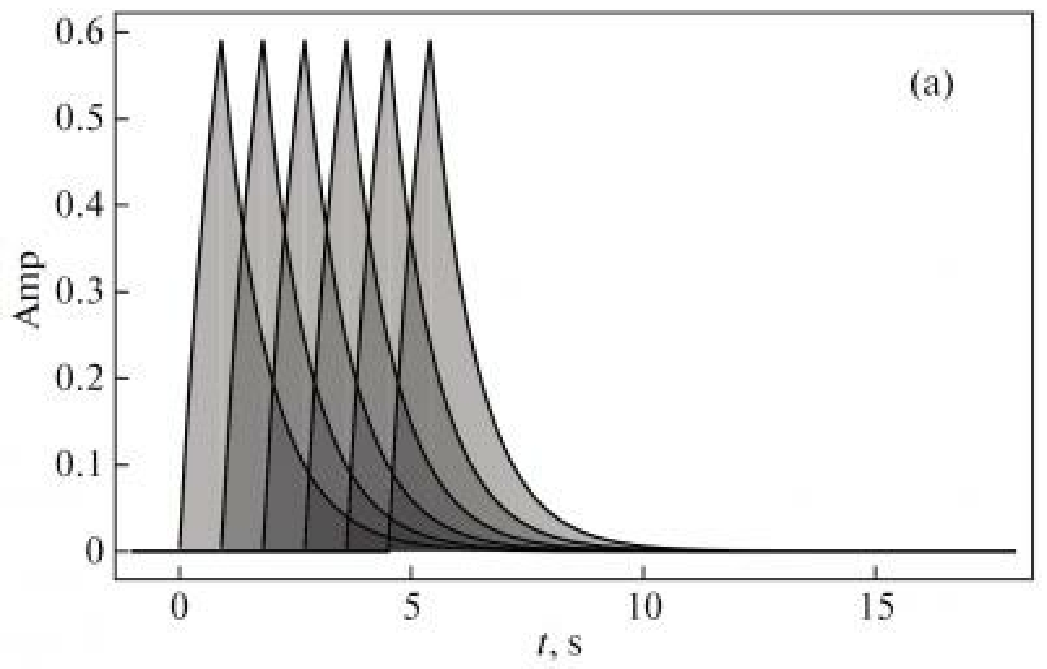}}
\end{minipage}
\hfill
\setcaptionmargin{1mm}
\begin{minipage}[h]{0.49\linewidth}
\center{\includegraphics[width=1\linewidth]{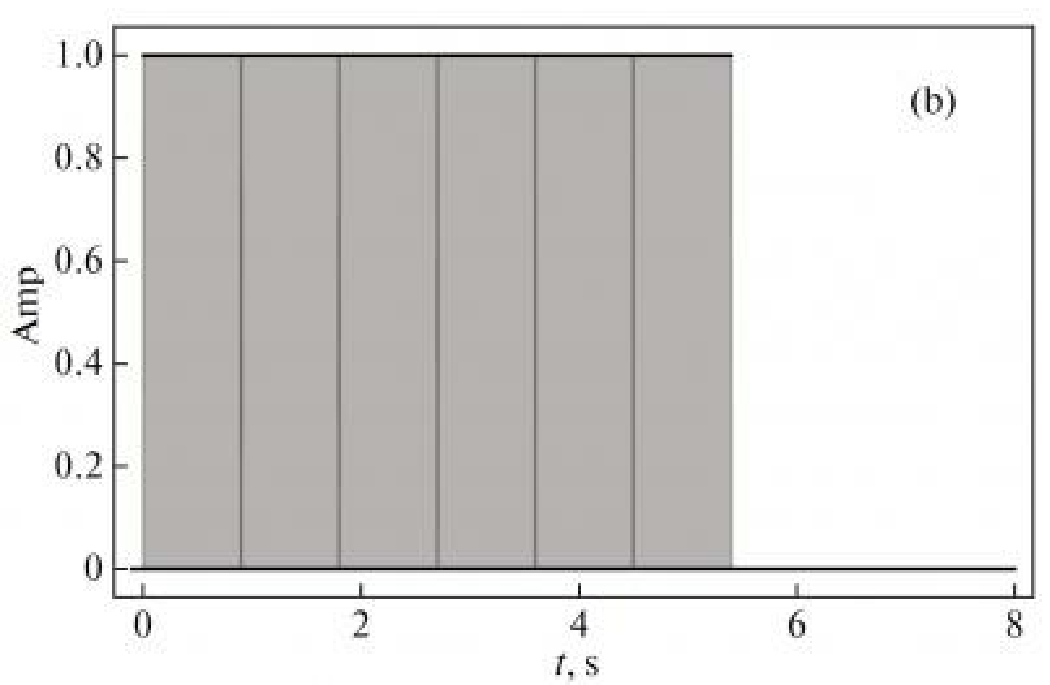}}
\end{minipage}
\label{ris:image1}
	\caption{(a) Simulations of the distortion of the shape of an impulse and the delay in each of the six frequency channels of the LSA. (b) $\Pi$- shaped function showing broadening of the pulse in the six frequency channels for $DM=360$ $pc \cdot ám^{-3}$. The
	broadening with a channel is $\Delta\tau_{DM=360}=0.8$ s.}
	\label{ris:fig2}
\end{figure}

\newpage
\begin{figure}[h!]
	\setcaptionmargin{1mm}
	\vbox{\includegraphics[width=0.9\linewidth]{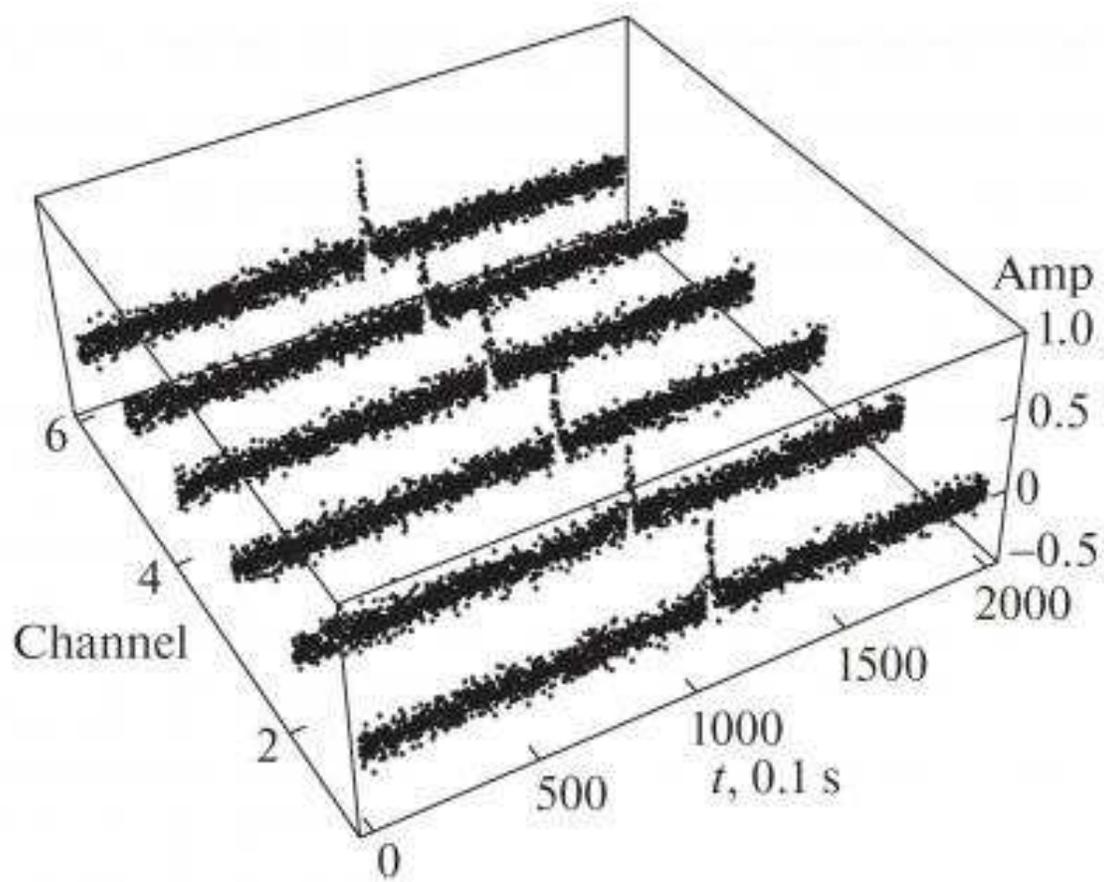}}
	\caption{Model for a received impulse against a white-noise
	background in the six frequency channels.}
	\label{ris:fig3}
\end{figure}

\newpage
\begin{figure}[h!]
\begin{minipage}[h]{0.85\linewidth}
\center{\includegraphics[width=0.8\linewidth]{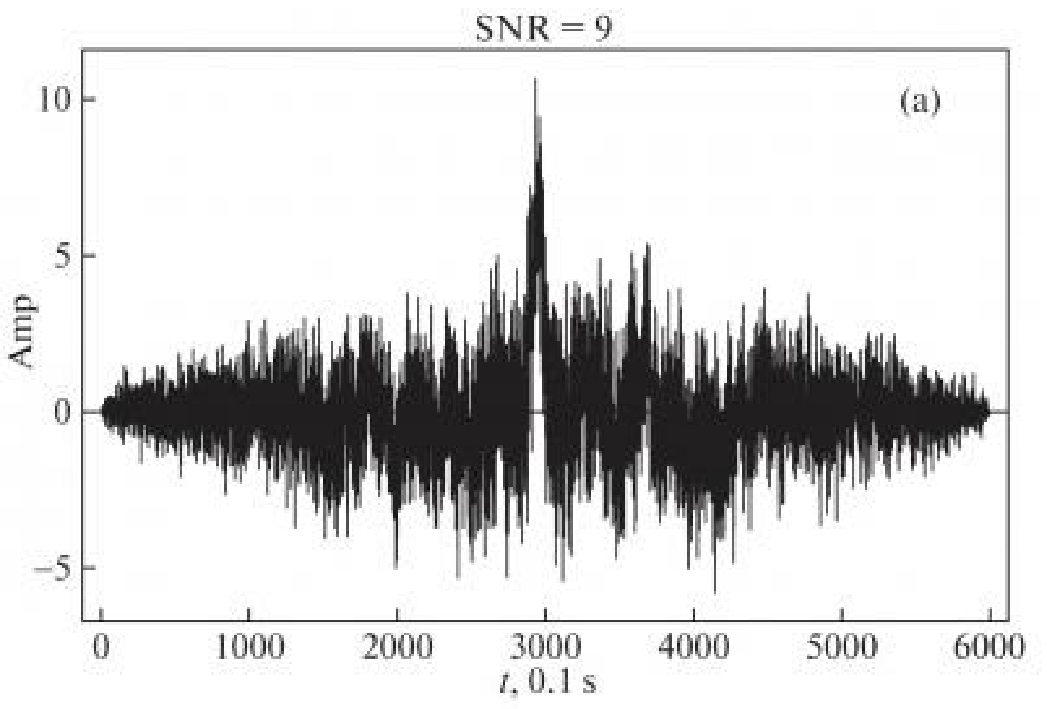}}
\end{minipage}
\hfill
\setcaptionmargin{1mm}
\begin{minipage}[h]{0.85\linewidth}
\center{\includegraphics[width=0.8\linewidth]{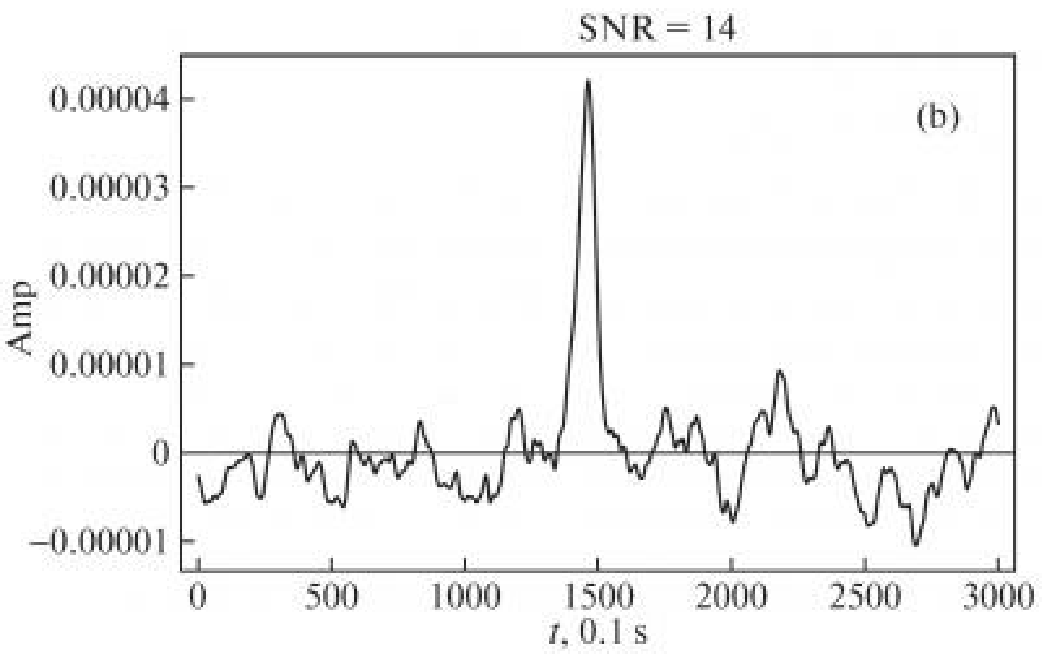}}
\end{minipage}
\vfill
\setcaptionmargin{1mm}
\begin{minipage}[h]{0.7\linewidth}
\center{\includegraphics[width=0.8\linewidth]{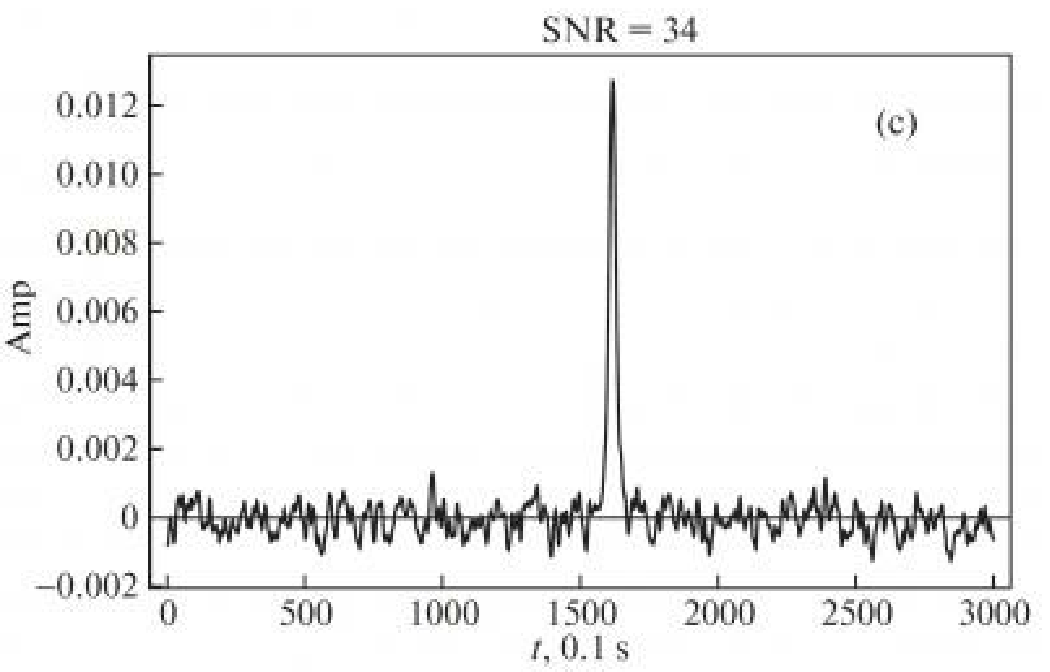}}
\end{minipage}
	\caption{Total impulse obtained as a result of (a) pair-wise cross-correlation of the six frequency channels without compensation for the dispersion measure, (b) cross - correlation with a template without compensation for the dispersion measure, and (c)
	cross - correlation with a template with compensation for the dispersion measure.}
	\label{ris:fig4}
\end{figure}

\newpage
\begin{figure}[h!]
\begin{minipage}[h]{0.8\linewidth}
\center{\includegraphics[width=1.1\linewidth]{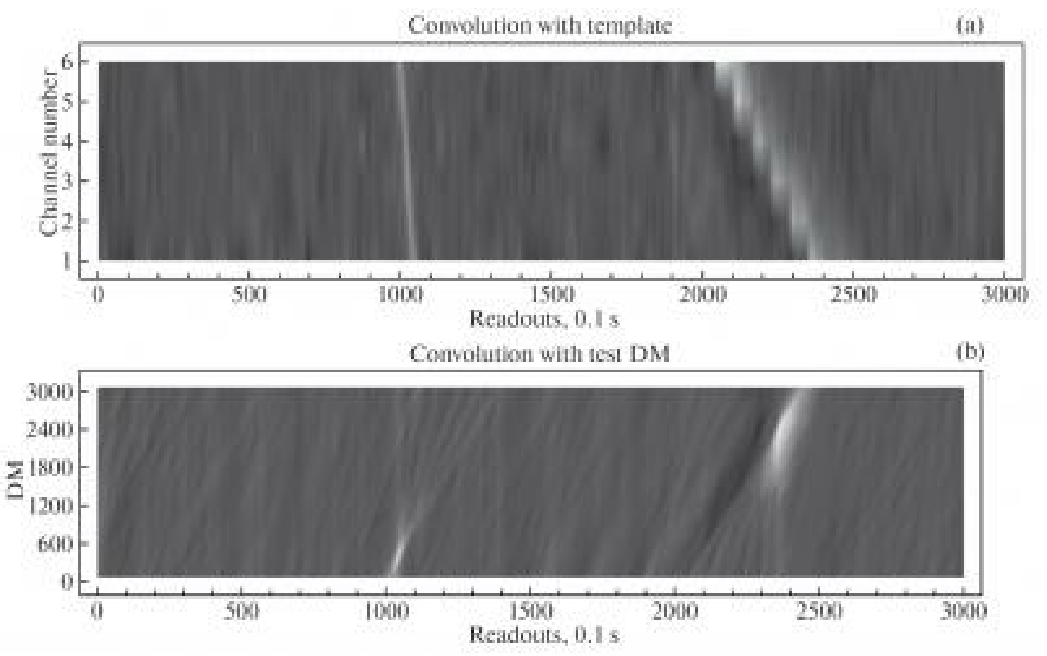}}
\end{minipage}
\vfill
\setcaptionmargin{1mm}
\begin{minipage}[h]{0.8\linewidth}
\center{\includegraphics[width=1.1\linewidth]{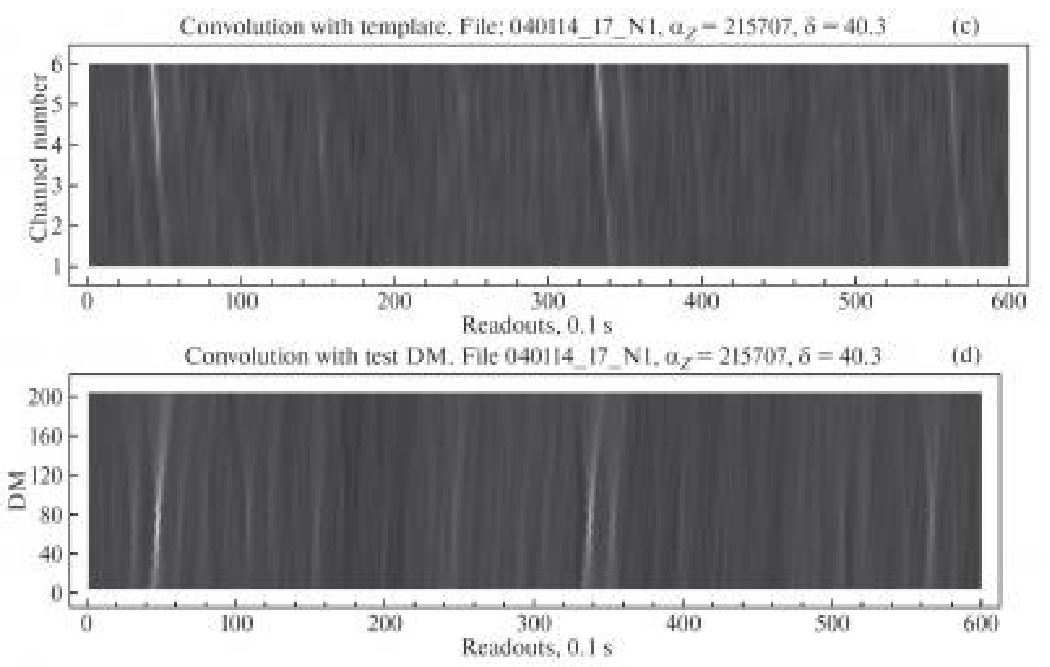}}
\end{minipage}
\label{ris:image1}
	\caption{ (a) Convolution with a template for a model impulse with dispersion measure 360 $pc/ám^3$ ¨ 2000 $pc/ám^3$. (b) Convolution of the impulses with a test dispersion measure. The desired dispersion measure is clearly visible in the plot as light sections whose brightness depends on the signal intensity. (c) Example of convolution of pulses from the pulsar B2154+40	with the template. We selected an interval with a duration of 60 s in which the individual pulses were clearly visible from an hour--long recording. (d) Result of convolving pulses from B2154+40 with a test dispersion measure. This pulsar has $DM = 78$ $pc/ám^3$. }
	\label{ris:fig2}
\end{figure}

\newpage
\begin{figure}[h!]
\setcaptionmargin{1mm}
\vbox{\includegraphics[width=1\linewidth]{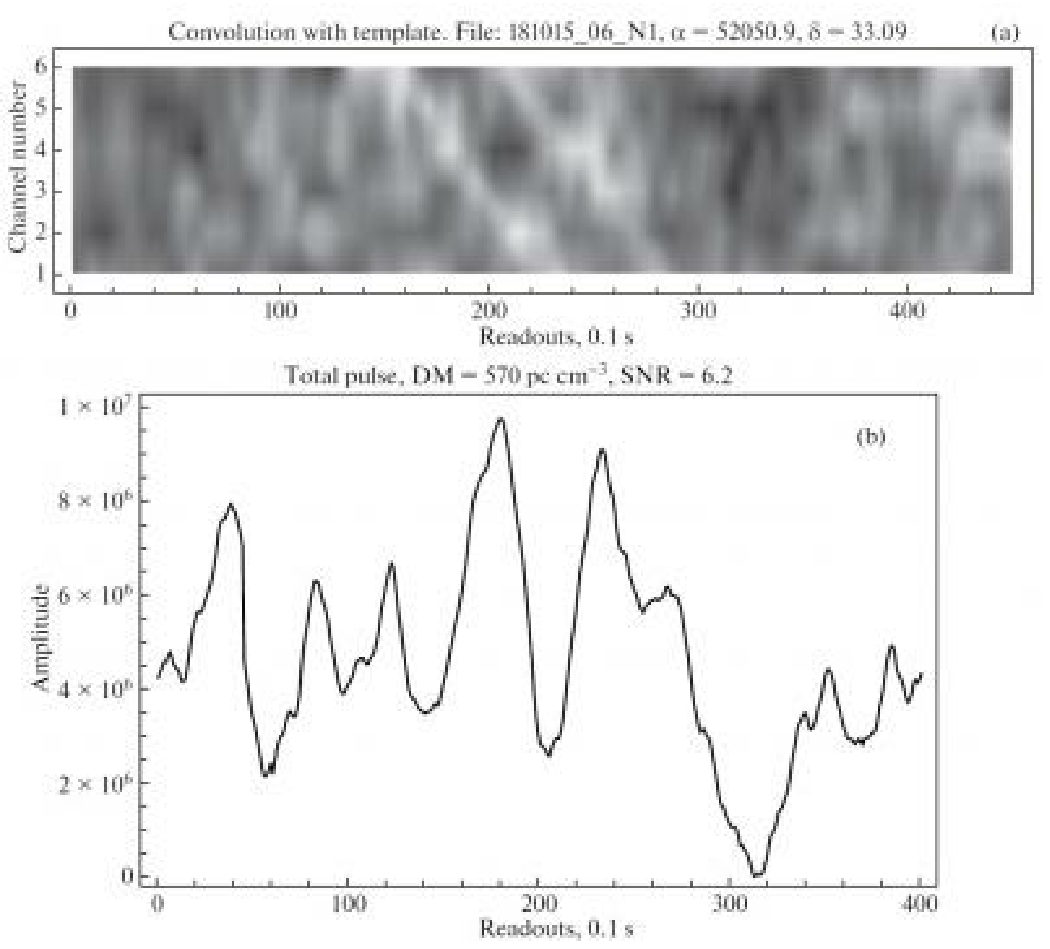}}
\label{ris:image1}
	\caption{(a) Dynamical spectrum of the impulse with $DM = 570$ $pc/ám^3$, detected on October 18, 2015. (b) Total pulse profile with $DM = 570$ $¯ª/á¬^3$. The peak flux density is 1.4 Jy, the Galactic coordinates are $b = 173.53^\circ$ and $l =$ -- 2.04$^\circ$, and the scattering is $t_s$ = 0.275 s.}
	\label{ris:fig2}
\end{figure}

\newpage
\begin{figure}[h!]
\setcaptionmargin{1mm}
\vbox{\includegraphics[width=1\linewidth]{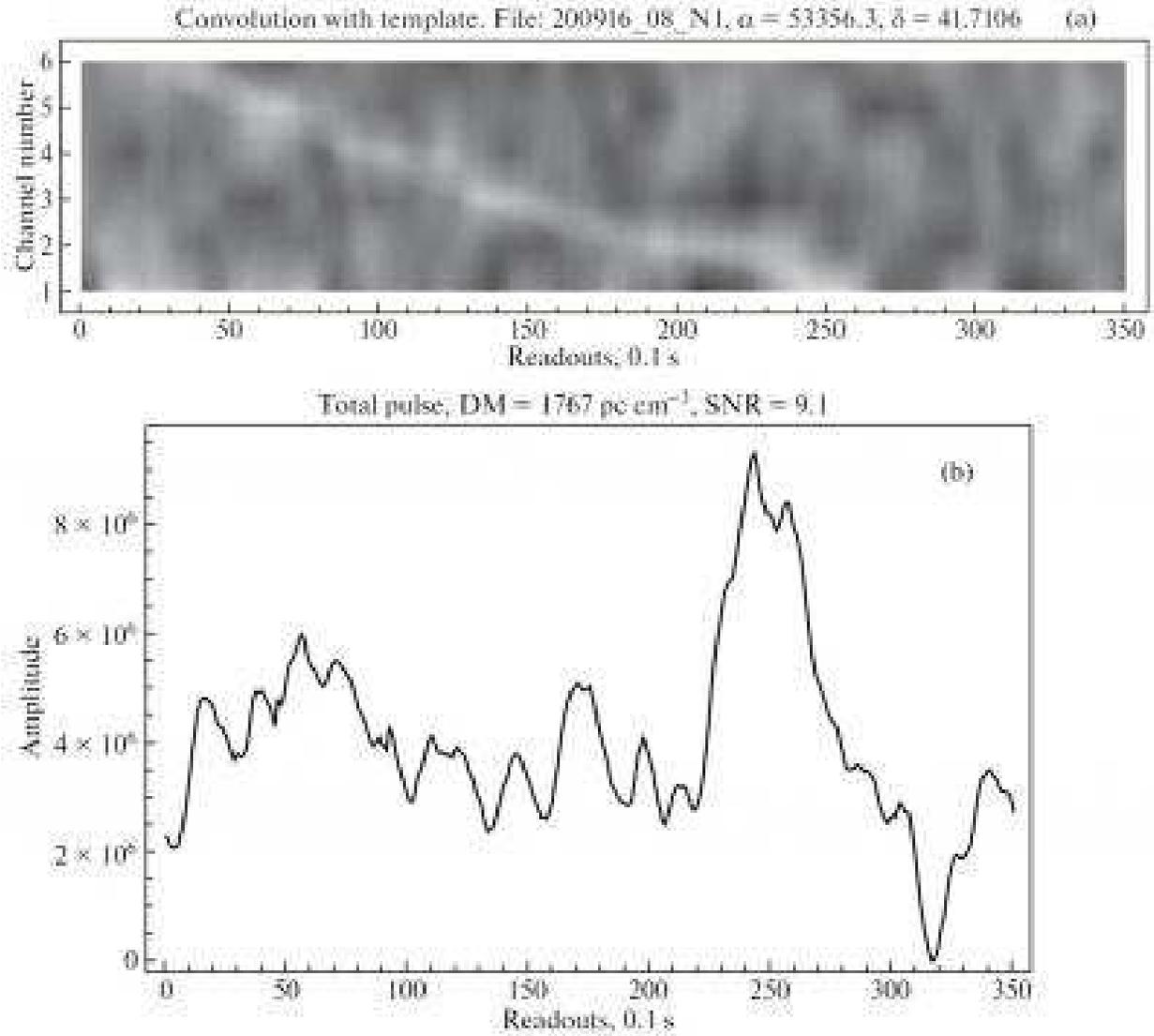}}
\label{ris:image1}
	\caption{(a) Dynamical spectrum of the impulse with $DM = 1767$ $pc/ám^3$, detected on September 20, 2016. (b) Total pulse	profile with $DM = 1767$ $pc/ám^3$. The peak flux density is 0.22 Jy, the Galactic coordinates are $b = 167.9^\circ$ and $l = +4.78^\circ$, and the scattering is $t_s$ = 4.78 s.}
	\label{ris:fig2}
\end{figure}

\newpage
\begin{figure}[h!]
\setcaptionmargin{1mm}
\vbox{\includegraphics[width=1\linewidth]{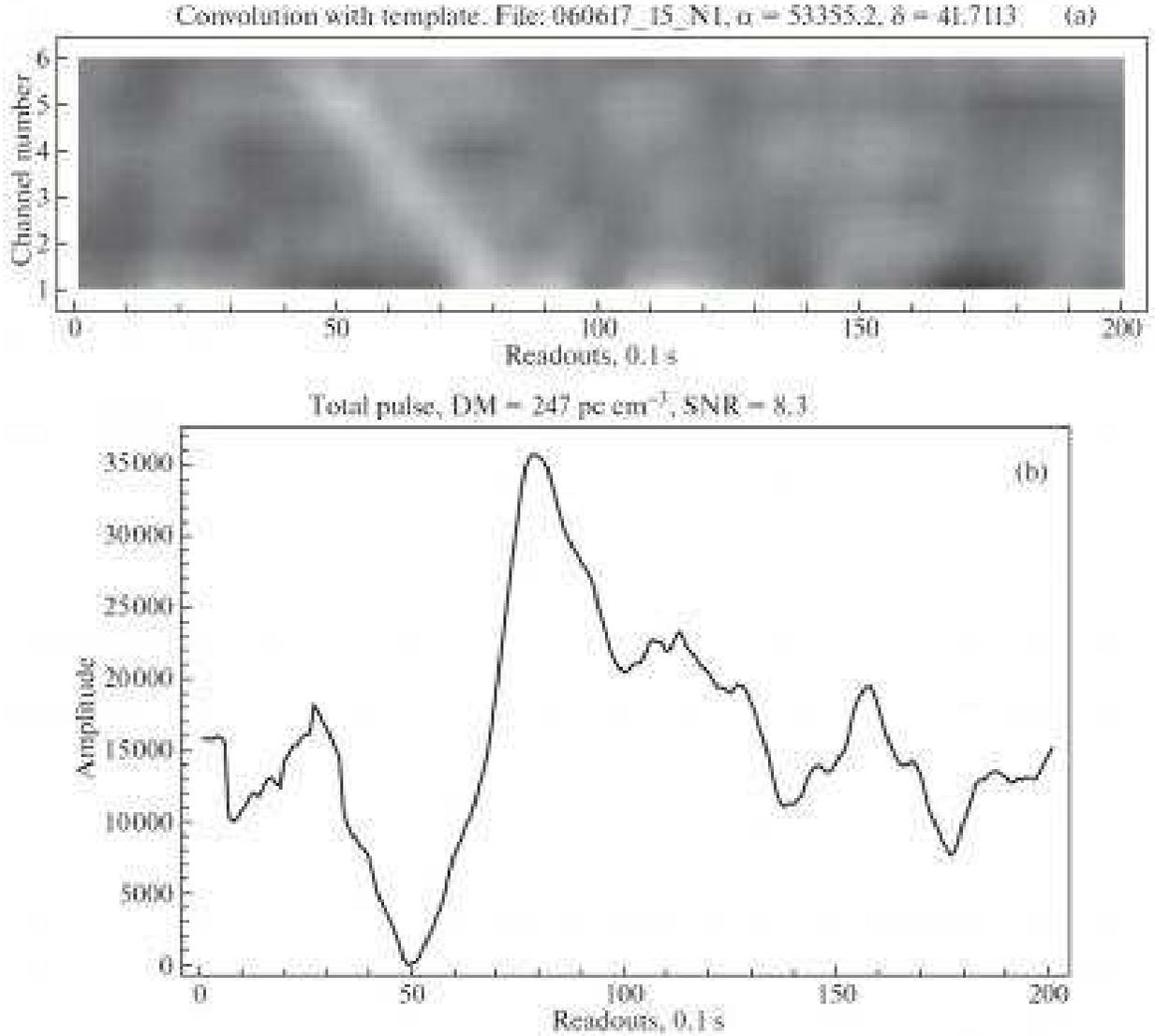}}
\label{ris:image1}
	\caption{(a) Dynamical spectrum of the impulse with $DM = 247$ $pc/ám^3$, detected on June 6, 2017. (b) Total pulse profile with $DM = 247$ $pc/ám^3$. The peak flux density is 0.54 Jy, the Galactic coordinates are $b = 167.9^\circ$ and $l = +4.78^\circ$, and the scattering is $t_s$ = 0.275 s.}
	\label{ris:fig2}
\end{figure}

\end{document}